# TetMaG-Guided Design and Operando Electron Holography Validation of Current-Induced Domain-Wall Motion in 3D Curved and Cornered Fe Nanobridges

Sameh Okasha[1*], Shriyar Tariq[1], Attila Kákay[2], Ryan Yang[2], Gregor Hlawacek[2], Akhil G. Nair[3], Rafal Dunin-Borkowski[3]

[1]SUPA, School of Physics and Astronomy, University of Glasgow (UoG), Glasgow G12 8QQ, UK.

[2]Institute for Ion Beam Physics and Materials Research, Helmholtz-Zentrum Dresden-Rossendorf (HZDR), Dresden, Germany

[3]Ernst Ruska-Centre for Microscopy and Spectroscopy with Electrons (ER-C), 52425 Jülich, Germany

[*]sameh.okasha@glasgow.ac.uk

**Abstract**:

Three-dimensional (3D) magnetic nanostructures offer new opportunities for controlling domain-wall (DW) configurations beyond the limitations of planar systems, providing promising architectures. However, the realization of reliable 3D magnetic devices requires precise control of geometry-dependent DW behaviour and quantitative experimental validation of the resulting magnetic states. Here, we combine TetMaG micromagnetic simulations, focused electron beam induced deposition (FEBID), and off-axis electron holography to investigate the influence of curvature and corner geometries on DW behaviour in 3D magnetic nanobridges.

TetMaG simulations predict fundamentally different magnetic properties for curved and cornered geometries. Cornered nanobridges act as preferential DW pinning sites, stabilizing localized magnetic configurations and enabling controlled switching between neighbouring pinning positions. While curved nanobridges promote gradual magnetization rotation, reduced pinning, and smoother DW motion. These optimized geometries were fabricated with high structural fidelity using FEBID and subsequently characterized by quantitative electron holography. Electron holography measurements revealed magnetic induction maps that matched the simulated magnetization configurations, providing direct experimental validation of the TetMaG predictions. Curved and cornered geometries exhibited distinct DW characteristics governed by their local structural features, demonstrating the critical role of geometry in tailoring magnetic behaviour in 3D systems. Operando current-biasing experiments further revealed current-induced DW motion, including the displacement of a tail-to-tail DW into a head-to-tail configuration within corner structures. Our results demonstrate a simulation-guided route for the design and experimental validation of geometry-controlled magnetic states and current-induced DW dynamics in 3D FEBID nanostructures. This approach provides a foundation for the development of next-generation 3D spintronic architectures towards 3D racetrack memory architectures.

## Introduction

Magnetic domain walls (MDWs) are fundamental spin textures that separate regions of different magnetization orientation and play a central role in emerging spintronic technologies.[1–14] Among these, racetrack memory has attracted considerable attention as a candidate for next-generation data storage owing to its potential for high storage density, non-volatility, and fast operation through current-induced DW manipulation. [13,15–19] In conventional racetrack architectures, magnetic nanowires are typically confined to planar geometries, limiting device density and restricting opportunities for exploiting 3D magnetic functionality.[20–24] The development of 3D magnetic nanostructures has opened new possibilities for controlling magnetic states through geometry[25,26]. Curvature, and corners, complex topologies introduce additional degrees of freedom that can influence magnetization reversal, DW pinning, and current-induced DW motion However, direct experimental validation of geometry-controlled DW behaviour in operando 3D magnetic nanostructures remains scarce. Recent advances in nanofabrication have enabled the realization of a variety of 3D magnetic systems, including curved nanowires, nanohelices, and freestanding magnetic architectures[2,27,28]. These studies have demonstrated that geometric confinement can significantly modify magnetic behaviour, offering new routes towards compact and multifunctional spintronic devices.[29] Focused electron beam induced deposition (FEBID) is particularly attractive for the fabrication of such structures because it enables direct-write growth of complex 3D magnetic geometries with nanometre-scale precision [30–33]. Combined with post-growth purification strategies and advanced nanofabrication techniques, FEBID provides a versatile platform for constructing magnetic nanostructures that are inaccessible using conventional lithographic approaches.[31,34,35] However, despite significant progress in magnetic nanofabrication, the experimental validation of 3D geometry-dependent magnetic states with current-induced DW behaviour remains limited.[26,36] Micromagnetic simulations provide a powerful framework for predicting magnetic configurations and DW dynamics in complex geometries. In particular, TetMaG [37]enables the simulation of magnetic structures with realistic 3D morphologies and has emerged as an effective tool for investigating geometry-dependent magnetic phenomena. Nevertheless, direct experimental verification of simulated magnetic states in 3D nanostructures remains challenging and requires quantitative magnetic imaging techniques with sufficient spatial resolution. Off-axis electron holography is uniquely suited to this task because it provides direct access to magnetic induction maps with nanometre-scale resolution.[38,39] The technique enables quantitative visualization of magnetic configurations and offers a powerful route for validating micromagnetic predictions in complex nanostructures. Furthermore, the combination of electron holography with operando electrical biasing provides opportunities to investigate current-induced magnetic phenomena directly within functioning devices.

In this work, we combine TetMaG micromagnetic simulations, FEBID nanofabrication, and off-axis electron holography to investigate geometry-dependent DW behaviour in curved and cornered 3D magnetic nanobridges. Cornered geometries act as preferential DW pinning sites, stabilizing localized magnetic configurations and enabling controlled switching between neighbouring pinning positions. In contrast, curved geometries promote gradual magnetization rotation, reduced pinning, and smoother DW propagation. TetMaG simulations are employed to optimize bridge geometries and identify configurations that DW pinning while preserving stable magnetic states. The resulting structures are fabricated in Fe using FEBID and characterised using quantitative electron holography. In addition, operando current-biasing experiments demonstrate current-induced DW motion within the fabricated nanobridges. These results establish a simulation-guided framework for the design, fabrication, and experimental validation of geometry-controlled magnetic states and current-driven DW dynamics in 3D magnetic nanostructures, providing a pathway towards 3D racetrack memory architectures

**Experimental:**

To establish a complete pathway from device design to magnetic characterization, we developed the workflow illustrated in Fig. 1. The process begins with TetMaG micromagnetic simulations, which are used to optimize the geometry of the magnetic nanobridges and predict their DW configurations and dynamics under current pluses. The optimized structures are subsequently fabricated using FEBID, enabling the realization of complex 3D geometries with high structural fidelity. Following fabrication, plasma purification and elemental analysis are performed to improve the metallic content and evaluate the material composition. The nanobridges are then integrated with an electrical biasing platform to enable operando current excitation. Finally, off-axis electron holography is employed to qualitatively map the magnetic induction and compare the experimentally observed magnetic states with the simulation predictions. The successful implementation of each stage establishes a complete simulation-guided workflow for investigating geometry-dependent magnetic behaviour and current-induced DW motion in 3D magnetic nanostructures.

**-Model-based growth**

3D Fe nanostructures were fabricated by focused electron beam induced deposition (FEBID) using a Thermo Fisher Helios 660 PFIB NanoLab at the Kelvin Nanocharacterisation Centre, University of Glasgow. The instrument was operated in conjunction with a computer-aided design (CAD)-compatible patterning software, enabling precise fabrication of complex three-dimensional geometries. The system provides a maximum imaging resolution of 6144 × 4096 pixels and a 64k patterning engine for high-precision beam control using $Fe_2(CO)_9$ precursor. A continuum growth model, F3ast[40], was developed for Fe under selected deposition conditions prior to fabricating the target nanostructures [33] that describes the influence of electron-beam-induced heating on FEBID growth in the desorption-limited regime, where precursor desorption becomes the dominant factor controlling deposition.

The deposition profile is approximated by a Gaussian distribution:

$$Deposition\ model = Gr * e^{\frac{-r^2}{2\sigma^2}} * e^{-KR_T}, \qquad (1)$$

where: *Gr* is the vertical growth rate at the center of the beam at base deposition temperature (nm/s) and is primarily determined by differences in gas pressure and precursor chemistry of Fe,

*K* is the thermal resistance scaling factor (dimensionless) and is mainly responsible for compensating for the heating effect during growth, σ is the standard deviation of the Gaussian Gr distribution, which determines the lateral spread of deposition around the beam center (nm), and

$R_T$ is the geometry-dependent factor accounting for total thermal resistance and beam-induced heating effects[41,42] (dimensionless). As beam-induced heating increases, precursor desorption becomes more pronounced, reducing local growth rates and potentially leading to structural instability. Consequently, the thermal response depends strongly on the geometry and connectivity of the structure, requiring geometry-specific optimization of deposition parameters to ensure stable growth and accurate shape reproduction.

Post-growth purification was carried out using a Fischione plasma cleaner operating with a 25% $O_2$/Ar mixture. The low-energy plasma (~12 eV, 13.56 MHz) was selected to preferentially remove carbonaceous species while preserving the structural integrity of the Fe nanostructures.

**-Electron holography characterization:**

Electron microscopy was carried out using an FEI Titan 80–300 TEM[43] equipped with a field-emission electron source and a CEOS CETCOR image-side spherical aberration corrector. Under optimized operating conditions, the instrument provides sub-ångström resolution with an information limit below 100 pm. Specimen positioning was achieved using a high-precision piezo-controlled stage, and images were recorded with a Gatan 2k × 2k CCD camera. Electron holograms were acquired under field-free conditions in Lorentz mode using an electrostatic biprism, which generated interference fringes. Magnetic phase reconstruction was performed using holograms recorded at reversal specimen tilt orientations to separate the electrostatic and magnetic phase contributions. The total phase shift $\emptyset(\mathrm{x})$ of the electron wave can be expressed as

$$\emptyset(x) = \emptyset_e + \emptyset_m$$
$$= C_E V(x)t(x) - \frac{\hbar}{et}\int B_\perp(x)t(x)dx \qquad (1)$$

where $\emptyset_e$ represents the electrostatic phase arising from the mean inner potential (MIP) of the specimen and $\emptyset_m$ is the magnetic phase associated with the magnetic vector potential. The incident electron beam direction z is perpendicular to x, $C_E$ is an interaction constant that takes a value of $6.53\times 10^6$ radV$^{-1}$ m$^{-1}$ at an accelerating voltage of 300 kV, e is elementary charge, t is local specimen thickness, $\hbar$ is reduced Planck constant, V is the electrostatic potential, and $B_\perp$ is the component of the magnetic vector potential along z.[38,39] The electrostatic phase was determined using the MIP method, while the magnetic phase was obtained by subtracting the electrostatic contribution from the reconstructed total phase. Magnetic induction maps can be obtained by adding magnetic phase image with contours which can be a direct representation of magnetic induction map as follow:

$$\frac{d\emptyset(X)}{dx} = -\frac{\hbar}{et}B_\perp(x) \qquad (2)$$

, where $\frac{d\emptyset(X)}{dx}$ is spatial gradient of the magnetic phase along the $x$-direction.

Operando current-biasing experiments were performed using a DENSsolutions Lightning in-situ TEM biasing holder equipped with a Lightning MEMS electrical biasing chip. The electrical pulses were generated using a pulse generator and applied to the nanobridges through the MEMS chip electrodes. The electrical setup was integrated with the TEM holder through the DENS interconnect system to enable synchronized current excitation during electron holography measurements.

Electron holograms were acquired before and after current excitation under identical Lorentz imaging conditions to investigate current-induced domain-wall motion while minimizing the influence of the microscope magnetic field.

**-TetMag Micromagnetic Simulation:**

Micromagnetic simulations were performed using TetMaG (Tetrahedral Micromagnetics)[37], a finite-element package specifically developed for the investigation of magnetic phenomena in complex 3D nanostructures. Unlike finite-difference methods based on regular cubic discretization, TetMaG employs tetrahedral meshes, enabling accurate representation of curved surfaces, sharp corners, and irregular geometries. This makes it particularly well suited for FEBID-fabricated structures, where geometry plays a critical role in determining magnetic states and DW behavior.

The simulations were based on the Landau–Lifshitz–Gilbert (LLG) equation, which describes the evolution of magnetization under the influence of exchange, magnetostatic, anisotropy, and external magnetic fields. To obtain equilibrium magnetic configurations, a damping parameter of $\alpha = 1.0$ was employed to accelerate convergence and suppress magnetization procession.

3D nanobridge geometries were generated as tetrahedral meshes using Gmsh and imported into TetMaG. The magnetic properties of Fe were defined using a saturation magnetization of Ms $= 1.7 \times 10^{6}$ A m$^{-1}$ and an exchange stiffness constant of $A = 2.1 \times 10^{-11}$ J m$^{-1}$. corresponding to bulk Fe at room temperature. The simulations yielded equilibrium magnetization distributions, DW configurations, and magnetization maps, which were directly compared with experimental electron holography measurements. This combined computational and experimental approach enabled quantitative assessment of geometry-dependent magnetic states, DW pinning, and current-driven magnetic behaviour in the investigated nanobridges.

The successful integration of the optimized nanobridges into the biasing platform establishes a key experimental capability required for quantitative studies of current-induced DW motion in 3D magnetic architectures.

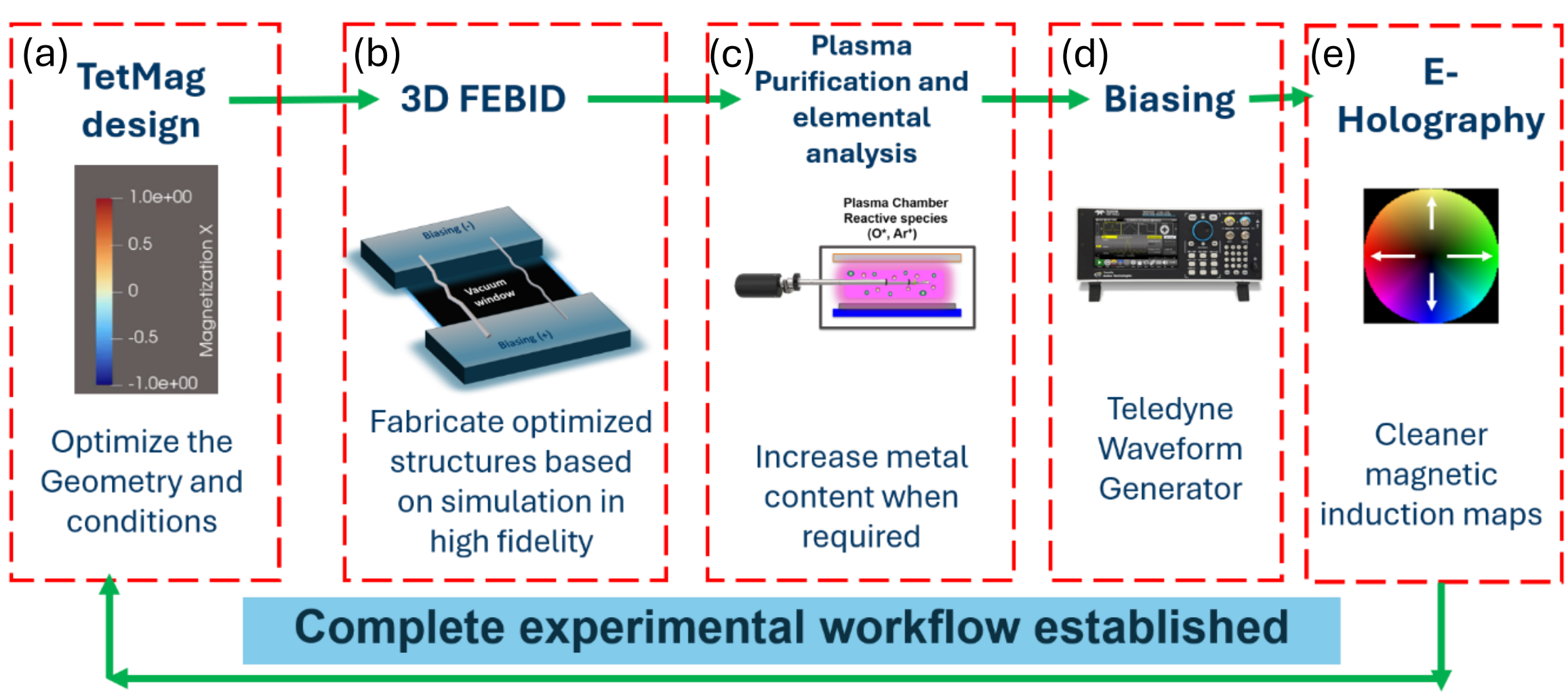

**Figure 1 | Schematic illustration of the experimental complete workflow established in this study.** (a)TetMaG micromagnetic simulations are first employed to optimize the geometry and operating conditions of curved and cornered magnetic nanobridges. (b)The optimized designs are subsequently fabricated with high fidelity using FEBID. (c) Post-growth plasma purification and elemental analysis are performed to enhance the metallic content and assess the material composition. (d)The fabricated structures are then integrated with an electrical biasing platform to enable current-induced DW motion. (e)Finally, off-axis electron holography is used to obtain qualitative magnetic induction maps and validate the simulated magnetic configurations. This integrated workflow provides a simulation-guided route for the design, fabrication, and experimental investigation of geometry-controlled magnetic states and current-induced DW dynamics in 3D magnetic nanobridges.

## Results and Discussion

To investigate the influence of geometry on magnetic behaviour, TetMaG micromagnetic simulations were performed on a series of curved and cornered nanobridge designs with different characteristic angles. Representative equilibrium magnetization states are shown in Fig. 2.

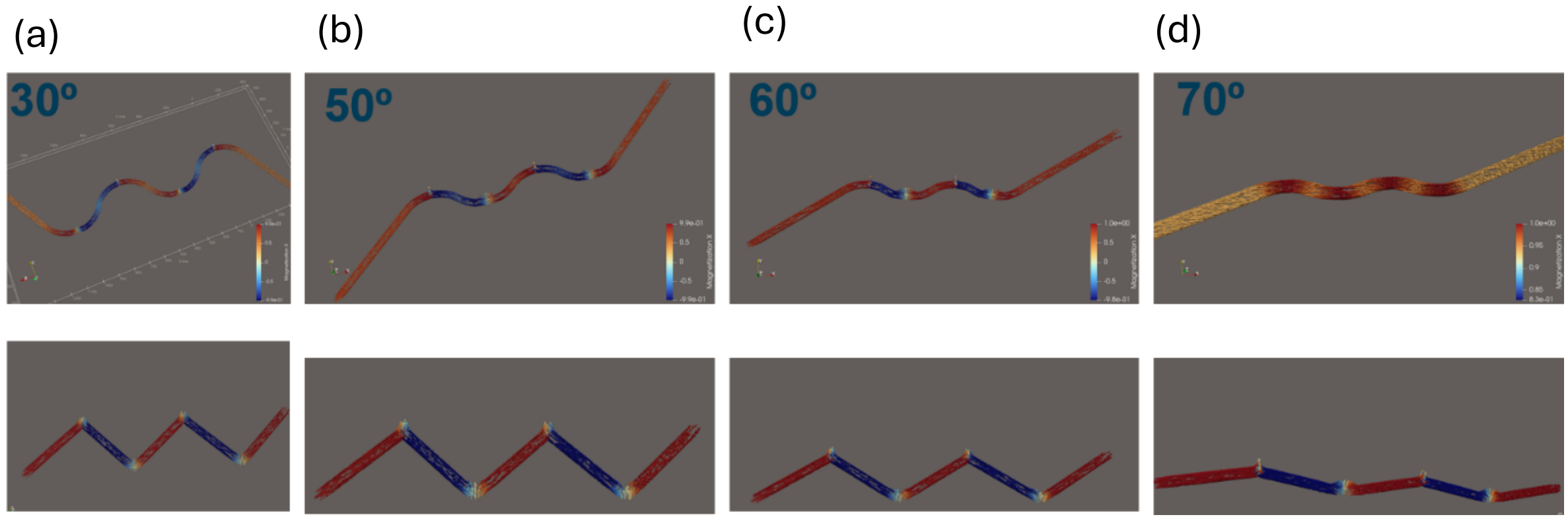

**Figure 2 | Geometry angle -dependent vs DW pinning tendency magnetic configurations in curved and cornered nanobridges predicted by TetMaG simulations.** Representative equilibrium magnetization configurations obtained from TetMaG simulations for Fe nanobridges with varying geometric angles. The upper row shows curved bridge geometries with characteristic angles of (a) 30°, (b)50°, (c)60°, and (d)70°, while the lower row presents corresponding cornered bridge geometries. The colour scale represents the magnetization component along the bridge axis, highlighting the formation and localisation of magnetic DWs within the structures. Significant variations in DW configuration and pinning behaviour are observed as the bridge geometry is modified. Curved geometries promote smoother magnetization transitions and reduced pinning, whereas cornered geometries generate localized magnetic inhomogeneities at the vertices that act as preferential pinning sites. These simulations were used to identify geometries that minimise DWs pinning while maintaining stable magnetic states for subsequent fabrication and experimental investigation.

The simulations reveal that the spatial distribution of magnetization is strongly governed by the local bridge geometry. In the curved structures, magnetization rotates gradually along the bridge contour, resulting in relatively smooth DW configurations and reduced localisation of magnetic charge. In contrast, cornered geometries introduce abrupt changes in direction that concentrate magnetic inhomogeneities at the vertices and promote DW pinning. As the bridge angle increases, systematic changes in the DW configuration are observed, indicating that geometric design can be used as an effective parameter for controlling magnetic states and DW stability. Based on these results, geometries exhibiting reduced pinning and favourable magnetic configurations were selected for FEBID fabrication and subsequent experimental validation by electron holography.

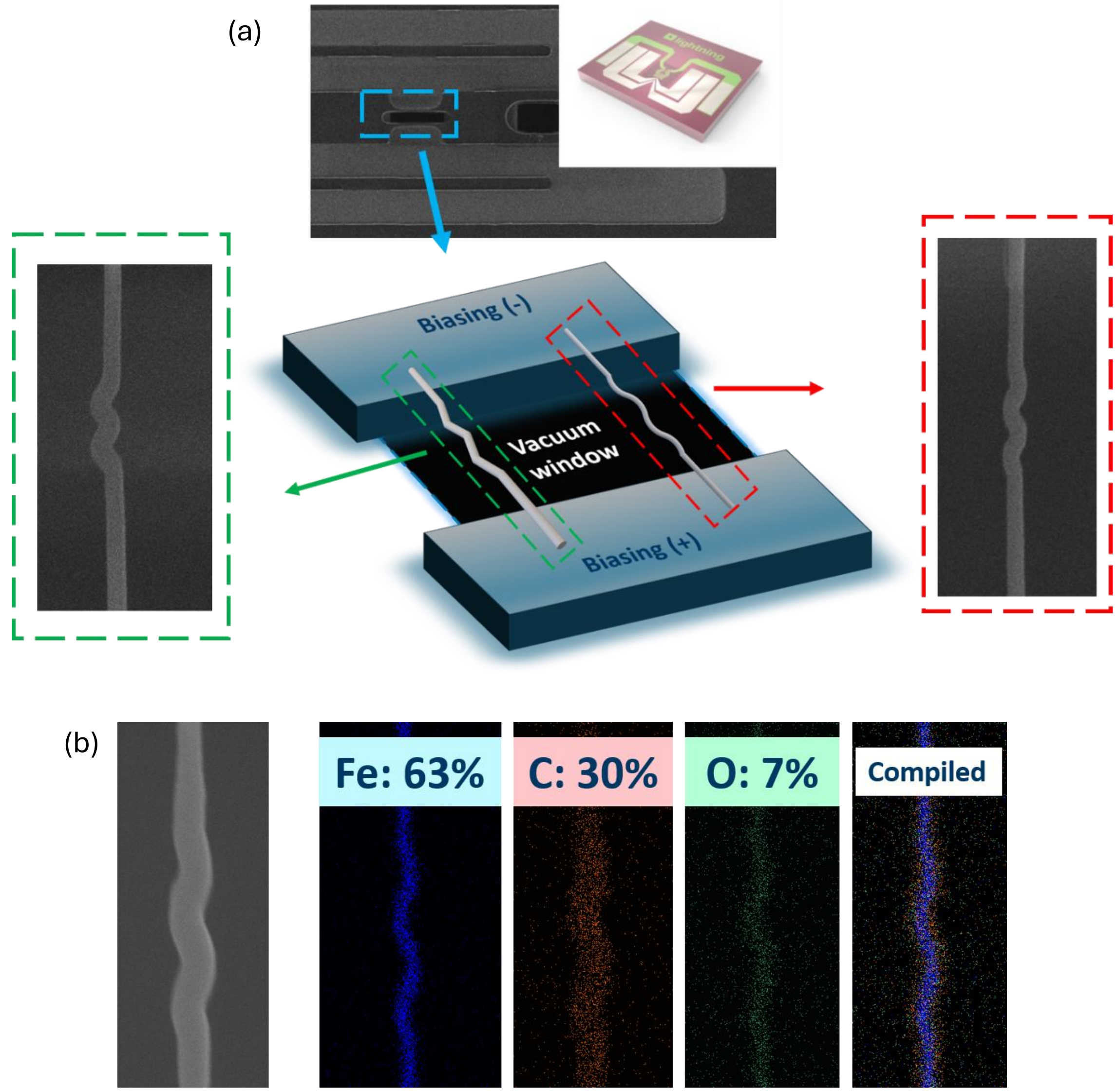

**Figure 3 | Integration of FEBID nanobridges into an operando electrical biasing platform.** (a) SEM image of the Dens-Chip system used for current-induced DW experiments. The FEBID nanobridge was fabricated across a vacuum window between two electrically connected biasing. The central schematic illustrates the device architecture and electrical biasing configuration. Enlarged SEM images of representative bridge geometries are shown on both sides, highlighting the fabrication fidelity and compatibility of the structures with operando electron holography measurements under applied electrical excitation. (b) compositional characterization of FEBID-grown Fe nanobridges consist of SEM-image of a representative curved Fe nanobridge fabricated by FEBID, Energy-dispersive X-ray spectroscopy (EDX) elemental maps showing the spatial distribution of Fe, C, and O within the structure. Quantitative compositional analysis reveals an Fe content of approximately 63 at. %, together with residual carbon and oxygen originating from the precursor decomposition process. Compiled elemental map demonstrating the homogeneous distribution of Fe throughout the nanobridge. The results confirm the successful fabrication of continuous magnetic nanostructures with sufficient metallic content for magnetic and electrical characterization. The nanobridges sustain current biasing up to 1 mA without structural failure

To investigate current-induced magnetic properties, the FEBID nanobridges were integrated into a Dens-Chip biasing system, as illustrated in Fig. 3. The structures were fabricated across a vacuum window connecting two biasing, enabling direct current operation through the magnetic bridge while maintaining compatibility with transmission electron microscopy and electron holography measurements. The device geometry was specifically designed to combine electrical functionality with magnetic imaging, allowing operando investigation of DW motion under applied current pulses. Following geometry optimisation by TetMaG simulations, the selected nanobridge designs were fabricated using FEBID as shown in SEM images of the resulting structures are shown in Fig. 3(a). The fabricated bridges closely reproduce the simulated geometries, demonstrating the capability of FEBID to realise complex 3D magnetic architectures with high structural fidelity.

To evaluate the material composition, SEM-EDX mapping was performed on the fabricated structures in-situ as deposited as shown in Fig. 3(b). The elemental maps reveal a homogeneous distribution of Fe throughout the nanobridge, with a measured Fe content of approximately 63 at.%. Residual carbon and oxygen were also detected, consistent with the incomplete decomposition of the organometallic precursor during FEBID growth. The relatively high Fe content obtained after growth optimisation provides sufficient magnetic contrast and electrical conductivity for subsequent operando experiments and electron holography characterisation. Plasma-based post growth purification has been utilized for short time to increase the Fe content % as well.

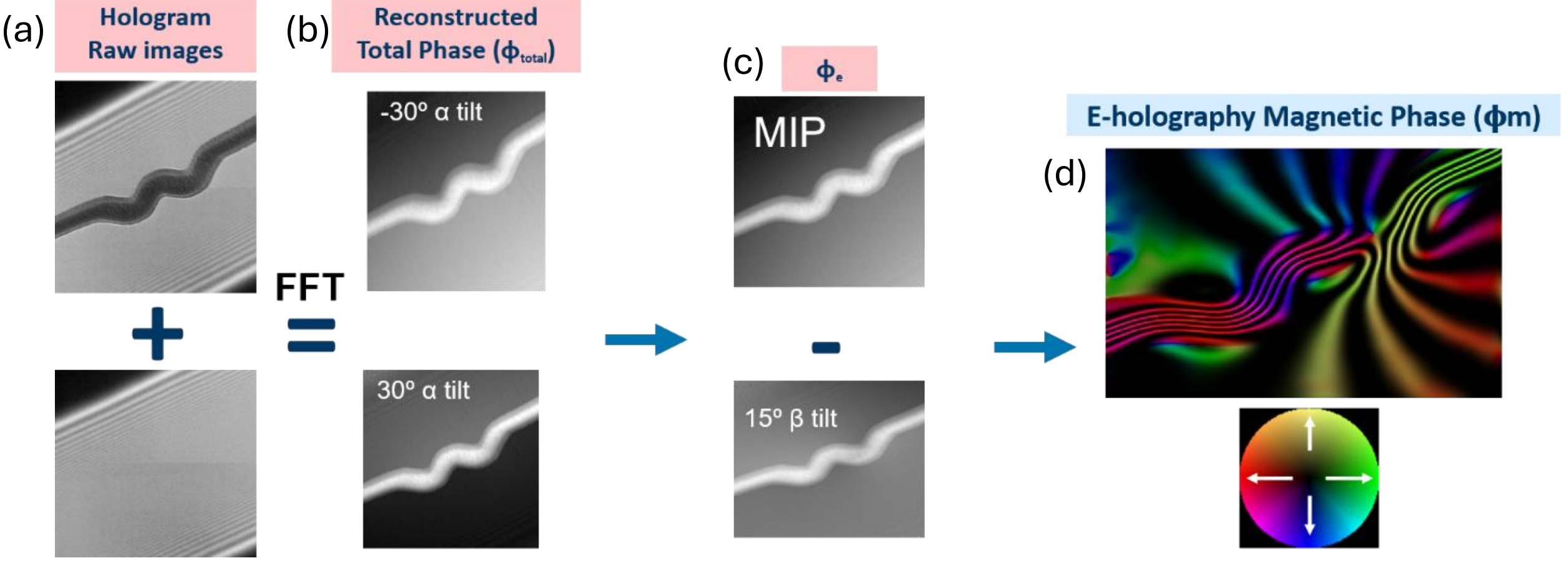

**Figure 4 | Electron holography reconstruction workflow for magnetic induction mapping in FEBID nanobridges.** (a) Raw off-axis electron holograms acquired from the magnetic nanobridge together with the corresponding reference holograms recorded under identical acquisition conditions. (b) Reconstructed total phase maps ($\phi_{total}$) obtained from holograms acquired at reversal α-tilt angles (−30° and +30°), enabling calculating the mean inner potential (MIP). (c) Electrostatic phase contribution ($\phi_e$) extracted using MIP method from holograms recorded at opposite β-tilt angles. (d) Magnetic phase map ($\phi_m$) obtained by subtracting the electrostatic contribution from the reconstructed total phase. The resulting magnetic phase contours directly reveal the magnetic flux distribution within and around the nanobridge providing a direct visualization of the magnetic DW in X-Y direction.

To quantitatively determine the magnetic configuration of the FEBID nanobridges, off-axis electron holography was employed following the reconstruction procedure illustrated in Fig. 4. Raw holograms were first acquired together with reference holograms under identical imaging conditions to remove instrumental contributions. The total phase shift, containing both electrostatic and magnetic information, was reconstructed from holograms recorded at reversal α-tilt (−30° and +30°). To isolate the magnetic contribution, the electrostatic phase was determined using the mean inner potential (MIP). Holograms acquired at 15 β-tilt were used to enable the DW visualization in X-Y direction. Subtraction of this electrostatic phase from the total reconstructed phase yielded the magnetic phase distribution (ϕm), which directly reflects the magnetic flux enclosed by the electron beam. The reconstructed magnetic phase map reveals the magnetic flux distribution associated with the domain structure of the nanobridge. From this phase information, magnetic induction maps were obtained, providing direct visualization of the magnetic configuration and DW localization within the 3D structure. The resulting induction maps exhibit strong agreement with the corresponding TetMaG simulations, confirming both the validity of the reconstruction procedure and the predictive capability of the simulation-guided design approach. The combination of off-axis electron holography and quantitative phase reconstruction provides a powerful methodology for investigating complex magnetic states in 3D nanostructures. Importantly, the ability to separate electrostatic and magnetic contributions enables direct comparison between experimentally measured magnetic induction and micromagnetic simulations.

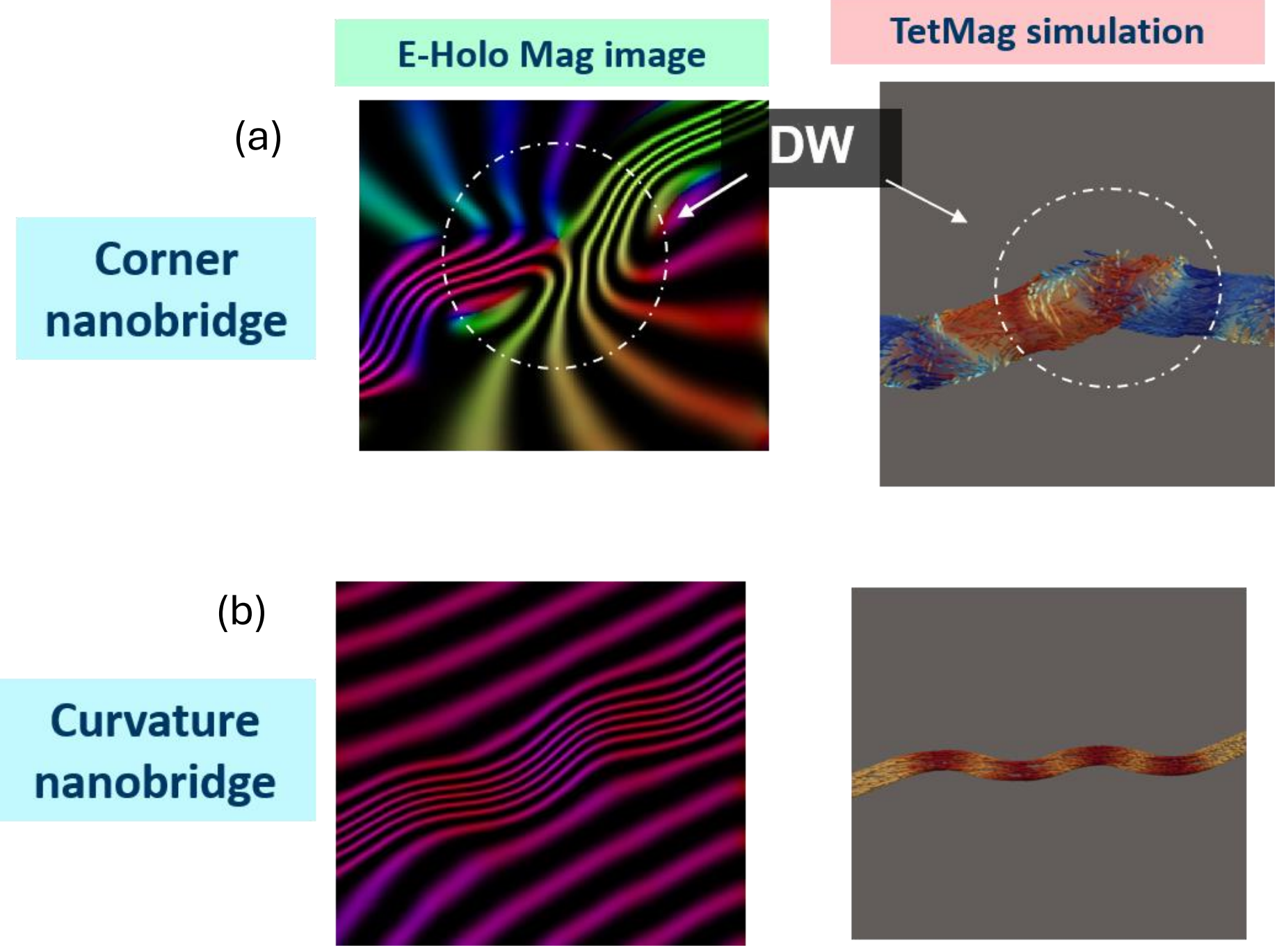

**Figure 5 | Electron holography validation of geometry-dependent DW configurations in 3D FEBID nanobridges.** Comparison between experimental magnetic induction maps obtained by off-axis electron holography and corresponding TetMaG micromagnetic simulations for cornered and curved Fe nanobridges. The left column shows reconstructed magnetic induction maps derived from electron holography in X-Y direction, while the right column presents the simulated magnetization configurations. For the cornered nanobridge (top row), both experiment and simulation reveal the formation of a localized DW at the geometrically confined region, highlighted by the dashed circles. For the curved nanobridge (bottom row), a smooth magnetization transition is observed along the bridge contour, consistent with the simulated magnetic configuration.

To experimentally validate the TetMaG predictions, the magnetic configurations of the fabricated nanobridges were investigated using off-axis electron holography. Figure 5 compares the reconstructed magnetic induction maps with the corresponding simulated magnetization distributions for representative cornered and curved bridge geometries. For the cornered nanobridge, both the experimental induction map and the simulation reveal a localized DW configuration at the geometrical constriction, demonstrating that the abrupt change in bridge direction acts as a preferential site for DW localization. The excellent agreement between the measured and simulated magnetic structures confirms the accuracy of the geometry-dependent micromagnetic model. In contrast, the curved nanobridge exhibits a gradual rotation of magnetization along the bridge contour, resulting in a smooth magnetic induction profile without strong localization of magnetic charge. The experimentally reconstructed induction map closely reproduces the predicted magnetic configuration, indicating that the curved geometry promotes continuous magnetization transitions and reduced pinning. The quantitative agreement between electron holography and TetMaG simulations provides direct experimental evidence that DW configurations in 3D FEBID nanostructures can be engineered through geometric design. These results establish electron holography as a powerful validation tool for simulation-guided development of 3D spintronic architectures. The correspondence between experiment and simulation demonstrates that DW localization and magnetic flux distribution can be programmed through geometrical design, providing a predictive route for engineering 3D magnetic devices.

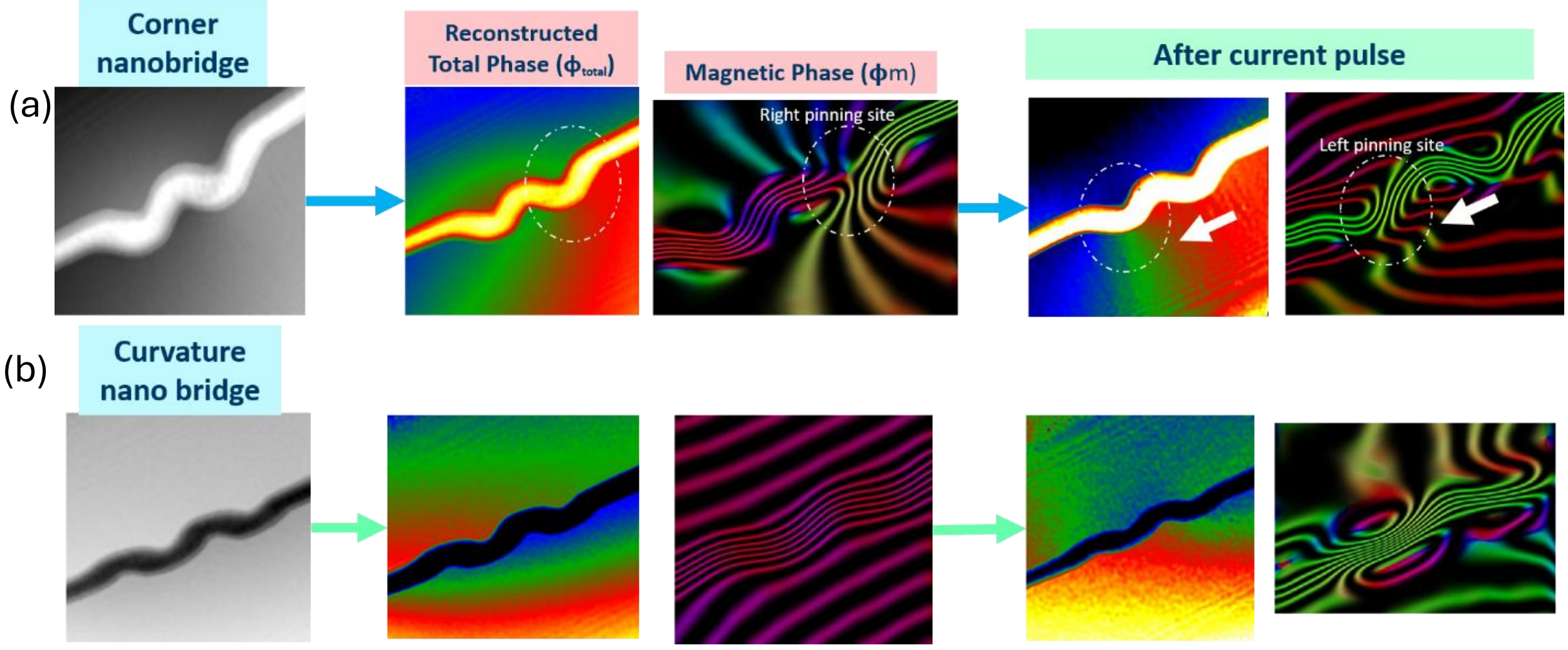

**Figure 6 | Experimental observation of current-induced DW motion in cornered and curved FEBID nanobridges.** Comparison of the magnetic configurations before and after the application of electrical current pulses in cornered and curved Fe nanobridges. For each geometry, reconstructed total phase maps and corresponding magnetic phase distributions obtained by off-axis electron holography are shown. In the cornered nanobridge (top row), the DW is initially localized at a geometrically defined pinning site and subsequently displaced following current flow, as indicated by the shift in the magnetic phase contours and the change in DW position between the right and left pinning sites. This corresponds to a transition from a tail-to-tail DW configuration to a head-to-tail DW configuration.In the curved nanobridge (bottom row), the magnetic phase distribution exhibits a smoother evolution under current flow, reflecting the reduced pinning potential associated with the continuous geometry. The observed changes in magnetic phase provide direct experimental evidence of current-induced DW motion and demonstrate the strong influence of geometry on DW dynamics in 3D magnetic nanostructures.

To investigate MDW motion under current-induction, operando electron holography measurements were performed before and after the application of electrical current pulses. Figure 6 compares the reconstructed magnetic configurations of cornered and curved Fe nanobridges and reveals clear geometry-dependent differences in DW motion dynamics. For the cornered nanobridge, the reconstructed magnetic phase maps show that the DW is preferentially localized at the corner as pinning site. Tail-to-tail DW is observed within the corner nanobridge. Detailed examination of the magnetic phase contours further reveals the formation of transverse-like DW containing vortex–antivortex magnetic structures. These features are evidenced by the presence of closed phase contours and the characteristic curling magnetic phase distribution surrounding the DW region.

Following the application of a current pulse, a clear displacement of the DW is observed, and evolves into a head-to-tail configuration accompanied by a redistribution of the magnetic phase contours. The DW migrates between adjacent corners of pinning sites, demonstrating that the geometrically defined corners act as controllable energy barriers governing DW motion. This behaviour is consistent with the TetMaG simulations, which predicted enhanced DW pinning at abrupt changes in bridge direction. In contrast, the curved nanobridge exhibits a more gradual magnetic phase evolution under identical current pulse conditions. The absence of sharp geometrical discontinuities reduces the local pinning potential, resulting in smoother DW propagation along the bridge contour. The reconstructed magnetic phase maps reveal a continuous redistribution of magnetic flux, indicating that curvature promotes more uniform DW motion compared with cornered geometries. These observations provide direct experimental evidence that current-induced DW motion can be engineered through geometrical design in 3D FEBID nanostructures. The results establish a direct link between nanobridge geometry, DW pinning properties, and current-induced dynamics. Importantly, the ability to manipulate DW between predefined pinning sites demonstrates a key operational principle required for racetrack-memory technologies and highlights the potential of simulation-guided 3D magnetic conduits for future spintronic applications.

## Conclusion

In this work, we established a simulation-guided framework for the design, fabrication, and validation of geometry-controlled magnetic nanostructures by combining TetMaG micromagnetic simulations, FEBID nanofabrication, and off-axis electron holography. TetMaG simulations identified optimized curved and cornered nanobridge geometries, which were subsequently fabricated with high fidelity and experimentally characterized. We demonstrate that geometry serves as an effective control parameter for DW properties in 3D magnetic nanostructures, with corners promoting strong DW pinning and localization, while curved geometries enable smoother DW propagation.

Electron holography revealed agreement with the simulated magnetic configurations, confirming the predictive capability of the simulation-guided approach. Furthermore, operando current-biasing experiments directly demonstrated current-induced DW motion, including the displacement of a tail-to-tail DW into a head-to-tail configuration and the formation of transverse-like vortex–antivortex DW structures. These experimentally observed magnetic textures closely matched the corresponding TetMaG predictions.

Our results demonstrate that DW behaviour can be engineered through geometrical design in 3D magnetic nanostructures, providing a pathway towards functional 3D spintronic devices and future racetrack memory architectures.

**Acknowledgements**:

This project was funded by RIANA, through the European Union as part of the Horizon Europe call HORIZON-INFRA-2023-SERV-01 under grant agreement number 101130652. Views and opinions expressed are, however, those of the author(s) only and do not necessarily reflect those of the European Union, and the European Union cannot be held responsible for them. RIANA provided access to the Ernst Ruska-Centre for Microscopy and Spectroscopy with Electrons (ER-C) and the Ion Beam Center (IBC) at Helmholtz-Zentrum Dresden-Rossendorf (HZDR), a member of the Helmholtz Association, where the experiments were carried out. The author acknowledges András Kovács, at Ernst Ruska-Centre for Microscopy and Spectroscopy with Electrons (ER-C), Germany, in addition to Trevor P. Almeida, and MCMP group at university for Glasgow, UK for their support for this RIANA-accepted project.